\documentclass{article}
\usepackage{amsmath}
\usepackage{amssymb}
\usepackage{graphicx}

\begin{document}

\mbox{  }
\\[0,5 cm]
\mbox{  }

\begin{center}
\LARGE 
\textbf{Third-order interference and}

\textbf{a principle of `quantumness'}
\\[1,0 cm]
\normalsize
Gerd Niestegge

\scriptsize
gerd.niestegge@web.de
\\[1 cm]
\normalsize
\end{center}
\normalfont \itshape Abstract. \normalfont 
Are there physical, probabilistic or information-theoretic 
principles which characterize the quantum probabilities and 
distinguish them from the classical case as well as 
from other probability theories, or which reveal why quantum 
mechanics requires its very special mathematical formalism? 
The paper identifies the fundamental absence of third-order 
interference as such a principle of `quantumness'. Considering 
three-slit experiments, the concept of third-order interference 
was originally introduced by Sorkin in 1994.
\\[0,5 cm]
\section{Introduction}

Are there physical, probabilistic or information-theoretic 
principles which 
specify the quantum probabilities and distinguish 
them from the classical case as well as from other probability 
theories, or which reveal why quantum mechanics requires its 
very special mathematical formalism? The relativity principle 
from which Einstein derived his theory is here considered a 
prototype by many physicists. 
The more it is tried to exploit 
the quantum peculiarities in modern quantum information theory, 
quantum computing and quantum cryptography, the more vital 
becomes the search for such principles of `quantumness'. 
A non-exhaustive selection of recent research papers 
with varying approaches are Refs. 
\cite{ref-Barn}, \cite{ref-Barr}, \cite{ref-Bru}, \cite{ref-Ari}, \cite{ref-Har}, 
and \cite{ref-Wil}; references to further work in this direction
can be found particularly in Ref. \cite{ref-Bru}.
The present paper introduces a new approach based on 
Sorkin's concept \cite{ref-Sor} of higher-order interference.
\\[0,3 cm]
Of course, the formalism of quantum theory is well-defined 
by a large system of mathematical axioms. It perfecty describes 
all the quantum peculiarities. However, it does not reveal 
their origin because these axioms have only a mathematical 
meaning, but no obvious physical or probabistic interpretation. 
The statistical interpretation is a later add-on motivated 
empirically, but not evident from the mathematical structure 
of the theory.
\\[0,3 cm]
A most typical quantum phenomenon is the wave-like interference 
exhibited in two-slit experiments with small physical particles. 
Considering experiments with three slits, Sorkin \cite{ref-Sor} introduced 
the concept of third-order interference and detected that 
standard quantum mechanics rules out third-order interference.
In a recent paper \cite{ref-Nie2}, the author could show that a probability theory 
where third-order interference does not occur must necessarily 
be very close to standard quantum mechanics.
\\[0,3 cm]
This results in the following  classification. A probability 
theory, where there is no interference at all, is classical. 
A theory including second-order interference, but ruling out 
third-order interference is quantum mechanics. The principle 
of `quantumness' thus becomes the presence of second-order 
interference combined with the absence of third-order 
interference. A theory where third-order interference is 
possible would go beyond quantum mechanics.
\\[0,3 cm]
The mathematics behind these findings has been elaborated 
in detail in Ref. \cite{ref-Nie2}.
The present paper paper shall outline the major result in a less 
mathematical way and focus on its role in a principle of `quantumness'.
After a brief overview of classical probabilities, quantum theory and a more 
general probabilistic framework in the next section, second- and 
third-order interference will be considered in the third 
section, before their role in a principle of `quantumness' 
will be studied in the fourth one.

\section{Probability theories}

In quantum mechanics, the measurable quantities of a physical 
system are re\-presented by observables. Most simple are those 
observables where only the two discrete values 0 and 1 are 
possible as measurement result; they are called \textit{events}. 
Mathematical probability theory usually starts with the 
identification of a structure for the events. Classically, 
this is a Boolean algebra. However, it is well-known that 
quantum mechanics requires a more general, not necessarily 
Boolean structure called \textit{quantum logic}.
This was pointed out by von Neumann and Birkhoff \cite{ref-Bir} 
already in the early days of quantum mechanics.
\\[0,3 cm]
An \textit{orthogonality} relation and a partial \textit{sum} operation + 
defined only for orthogonal events are available on a 
quantum logic. Orthogonality means that the events exclude 
each other and, in the classical case, it is the same as 
disjointness. 
Standard quantum mechanics uses a very special type of quantum logic; 
it consists of the self-adjoint projections on a 
Hilbert space or, more generally, of the self-adjoint 
projections in a von Neumann algebra.
\\[0,3 cm]
The \textit{states} on a quantum logic are the analogue of the 
probability measures in classical probability theory, 
and \textit{conditional probabilities} can be defined similar to 
their classical prototype \cite{ref-Nie2}. However, there are many 
quantum logics where no states or no conditional 
probabilities exist, or where the conditional probabilities 
are ambiguous. Therefore, only those quantum logics where 
sufficiently many states and unique conditional 
probabilities exist can be considered a satisfying 
framework for general probabilistic theories.
\\[0,3 cm]
A state $ \mu $ allocates a probability $ \mu (e) $ to 
each event $ e $ in such a way that 
$ \mu (e_1 + e_2) = \mu (e_1) + \mu (e_2) $ 
holds for any two orthogonal events $ e_1 $ and $ e_2 $. 
The conditional probability of an event $ f $ under 
another event $ e $ in the state $ \mu $ is denoted 
by $ \mu(f \mid e) $; this is the updated probability of the 
event $ f $ after the event $ e $ has been the outcome of
a first measurement. 
\\[0,3 cm]
With two successive measurements, the 
probability that the first one provides the result 
$ e $ and the second one then the result $ f $ is 
the product $ \mu(f \mid e) \, \mu (e) $.
In the classical case, the system of events forms 
a Boolean algebra and this probability becomes 
$ \mu(f \mid e) \, \mu (e)  = \mu(f \cap e) $, 
which is additive in $f$ as well as in $e$. 
In the general case, it remains additive only in $f$,
but not in $e$. This is the origin of quantum interference
which shall be considered in the following section.

\section{Interference} 

For a pair of orthogonal events $ e_1 $ and $ e_2 $, 
a further event $f$ and a state $ \mu $, the
following mathematical term $I_2$ shall be studied: 
$$
I_2 := \mu(f \mid e_1 + e_2) \,  \mu(e_1 + e_2) -  \mu(f \mid e_1) \, \mu(e_1) - \mu(f \mid e_2) \, \mu(e_2)  
$$
For classical probabilities, the identity $ I_2 = 0 $ 
holds, but not for 
quantum mechanics. Many quantum peculiarities can directly 
be traced back to the fact that $I_2 = 0$ is not valid. 
This is the essence of quantum interference which is exhibited 
e.g., in the two-slit experiments with small physical 
particle like electrons, photons and others, or in experiments measuring 
the spin of electrons and photons twice along differently oriented spatial axes. 
\\[0,3 cm]
Sorkin \cite{ref-Sor} introduced the concept of third-order interference. 
For a triple of orthogonal events $ e_1 $, $ e_2 $ and $ e_3 $, 
a further event $f$ and a state $ \mu $, he defined the following 
mathematical term $I_3$:
\\
$$
\begin{array}{rcl}
  I_3 & := & \mu(f \mid e_1 + e_2 +e_3) \, \mu(e_1 + e_2 + e_3) \\
    &   &   \\
    &   & - \mu(f \mid e_1 + e_2) \, \mu(e_1 + e_2) \\
    &   &   \\
    &   & - \mu(f \mid e_1 + e_3) \, \mu(e_1 + e_3) \\
    &   &   \\
    &   & - \mu(f \mid e_2 + e_3) \, \mu(e_2 + e_3) \\
    &   &   \\
    &   & +  \mu(f \mid e_1) \, \mu(e_1) + \mu(f \mid e_2) \, \mu(e_2) + \mu(f \mid e_3) \, \mu(e_3) \\
\end{array}
$$
\\[1,0 cm]
Sorkin's original definition refers to probability measures 
on `sets of histories'. With the use of conditional probabilities,
it gets the above shape, which was seen by Ududec, Barnum and 
Emerson \cite{ref-Udu} who adapted it into an operational probabilistic 
framework.
\\[0,3 cm]
Classical probabilities satisfy the identity $ I_3 = 0 $; this follows 
from $ I_2 = 0 $. Sorkin detected that $ I_3 = 0 $ also holds in 
quantum mechanics.
The identity $ I_3 = 0 $ can be rewritten in the following form:
\\[0,3 cm]
$
\mu(f \mid e_1 + e_2 +e_3) \, \mu(e_1 + e_2 + e_3) - \mu(f \mid e_1) \, \mu(e_1) - \mu(f \mid e_2) \, \mu(e_2) - \mu(f \mid e_3) \, \mu(e_3)
$
\begin{flushright}
$
= \mu(f \mid e_1 + e_2) \, \mu(e_1 + e_2) - \mu(f \mid e_1) \, \mu(e_1) + \mu(f \mid e_2) \, \mu(e_2)
$
\\[0,3 cm]
$
+ \, \mu(f \mid e_1 + e_3) \, \mu(e_1 + e_3) - \mu(f \mid e_1) \, \mu(e_1) + \mu(f \mid e_3) \, \mu(e_3)
$
\\[0,3 cm]
$
+ \, \mu(f \mid e_2 + e_3) \, \mu(e_2 + e_3) - \mu(f \mid e_2) \, \mu(e_2) + \mu(f \mid e_3) \, \mu(e_3)
$
\end{flushright}
The left-hand side of this equation is a measure for the interference 
involved in the case with three mutually exclusive events $e_1$, $e_2$, and $e_3$. It is identical 
with the sum of the measures for the pair interferences when the three different pairs 
are considered which can be built from the three events. Therefore, 
$ I_3 = 0 $ means that interference with three mutually exclusive events
does not provide anything new compared to the cases with only two exclusive events.
However, $ I_3 \neq 0 $ would mean that a fundamentally new form of interference 
exists in addition to the pair interferences.
\\[0,3 cm]
While the familiar two slit-experiment involves only two possible paths, 
three different paths are available for the particle in a three-slit 
configuration. In this case, the identity  $I_3 = 0$ means that quantum interference 
is limited to pairs of paths and that quantum mechanics does not exhibit 
a new form of interference involving path triples. The interference pattern
observed with three open slits is a simple combination of the patterns observed 
in the six different cases with only one or two among the three available slits open,
which could be confirmed in a recent experimental test with photons \cite{ref-Sin}.  
\\[0,3 cm]
As well as $I_2$, the term $I_3$ defines a very general 
concept and is not restricted to multiple-slit experiments.
The new type of interference which is present whenever $ I_3 \neq 0 $ 
holds is called \textit{third-order interference}, and the one present
whenever $ I_2 \neq 0 $ holds is called \textit{second-order interference}.
\newpage
\section{A principle of `quantumness'}

The absence of second-order interference ($I_2 = 0$) characterizes 
the classical probabilities. The presence of second-order 
interference ($ I_2 \neq 0 $) is typical of the quantum probabilities, 
but third order-interference is not possible in quantum theory. 
To what extent does now the absence of third-order interference 
($I_3 = 0$) characterize the quantum probabilities? Can there be 
other probability theories with $I_3 = 0$, or is quantum mechanics 
the only one?
\\[0,3 cm]
In Ref. \cite{ref-Nie2}, it could be shown that that a probability theory where 
third-order interference does not occur must necessarily be 
very close to standard quantum mechanics. The events must then be 
projections in a Jordan algebra.
This is the major result (Theorem 11.1) of Ref. \cite{ref-Nie2}.
Only the exceptional 
Jordan algebras provide examples with $I_3 = 0$ which are not covered 
by standard quantum mechanics (i.e., do not have a representation 
as operators on a Hilbert space). 
This results in the classification of the different probability theories
which is depicted in Figure 1.
\\[0,3 cm]
The combination of the presence of second-order interference ($ I_2 \neq 0 $) 
with the absence of third-order interference ($I_3 = 0$) can therefore be 
identified as an essential principle of `quantumness'. 
The presence of second-order interference ($ I_2 \neq 0 $) distinguishes 
quantum theory from the classical case. It is the reason 
why classical probability theory cannot cover the quantum 
probabilities. The fundamental absence of third-order
interference ($I_3 = 0$) entails the very special mathematical 
formalism of quantum mechanics. It defines what quantum 
theory is, leaving only little room beyond standard 
quantum mechanics. A further still unknown principle 
might do the rest and rule out the exceptional Jordan 
algebras.
\begin{figure}[h]
	\centering
		\includegraphics[width=0.70\textwidth]{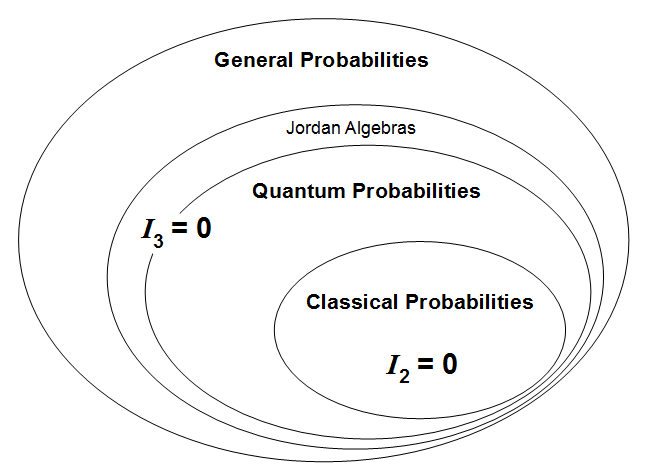}
	\caption{Classification of probability theories}
	\label{fig:Figure}
\end{figure}

\newpage
\section {Remarks}
The above results and their mathematical proofs have been
elaborated in detail in Ref. \cite{ref-Nie2}.
Without going too far into the mathematical details, 
some assumptions which they are based 
upon shall be presented here. The general probabilistic 
framework is given by those quantum logics where 
sufficiently many states and unique conditional 
probabilities exist. Beyond that, three further 
assumptions are needed.
\\[0,3 cm]
The first assumption is that real-valued (i.e., not 
necessarily positive) generalized states   
satisfy a Hahn-Jordan decomposition property similar 
to the classical real-valued measures.
\\[0,3 cm]
A quantum logic with sufficiently many states can always 
be embedded in an ordered linear space. In Ref. \cite{ref-Nie2}, 
it was shown that the absence of 
third-order interference ($I_3 = 0$) is equivalent to the 
existence of a product in this ordered linear space.
Generally, this product is neither commutative 
nor associative, and the second assumption is that each 
element of the ordered linear space generates an associative subalgebra. The 
third one is that the square of an element is positive. 
\\[0,3 cm]
The elements of the ordered linear space are candidates for observables,
for which the last two assumptions are quite natural 
postulates. The first assumption 
is a mathematical technical requirement.
\\[0,3 cm]
Only the exceptional Jordan algebras satisfy these assumptions
and the identity $I_3 = 0$ without being included 
in standard quantum mechanics.
There are not too many of them and, in a certain sense, there is
only one; it is formed by the self-adjoint $3 \times 3$-matrices  
the entries of which are octonions (see Ref. \cite{ref-Han}).

\section {Conclusions}

The concept of third-order interference is a quite natural 
extension of the second-order interference which is so 
typical of quantum mechanics, but surprisingly third-order 
interference is ruled out by quantum theory. Quantum interference 
involves only pairs, but no triples of mutually exclusive 
alternatives.
\\[0,3 cm]
In the present paper, the absence of third-order interference (i.e., $I_3 =0$) has been 
identified as a fundamental principle of `quantumness' which entails the 
very special structure of quantum theory. It defines the theory, 
leaving only little room beyond standard quantum mechanics.
\\[3,0 cm]
A violation of the identity $I_3 = 0$ has never been detected, 
and a recent experimental test \cite{ref-Sin} has confirmed it.
One can therefore still assume that quantum mechanics is universally 
valid in nature. However, its universal validity
is sometimes questioned because a successful unification 
with relativity theory and a satisfying quantum gravity theory 
are still missing. This might require a new theory going 
beyond standard quantum mechanics and possibly exhibiting third-order
interference.

\end{document}